# Thermal Nanoquakes: Terahertz Frequency Surface Rayleigh Waves in Diamond Nanocrystals

Caleb Stamper,[1] Matteo Baggioli,[2,3] Pablo Galaviz,[4] Roger A. Lewis,[5] Kirrily C. Rule,[4,5] Ablikim Bake,[1,6] Kyle A. Portwin,[1] Sha Jin,[2,3] Xue Fan,[7,8] Dehong Yu,[4*] and David L. Cortie[4,5*]

[1]*Institute for Superconducting and Electronic Materials, University of Wollongong, Wollongong, NSW 2500, Australia.*

[2]*School of Physics and Astronomy, Shanghai Jiao Tong University, Shanghai 200240, China*

[3]*Wilczek Quantum Center, Shanghai Jiao Tong University, Shanghai 200240, China*

[4]*Australian Nuclear Science and Technology Organisation, Lucas Heights, NSW 2234, Australia.*

[5]*School of Physics, University of Wollongong, Wollongong, NSW 2500, Australia.*

[6]*Institute for Glycomics, Griffith University, Gold Coast, QLD 4215, Australia.*

[7]*Shanghai National Center for Applied Mathematics, Shanghai Jiao Tong University, Shanghai 200240, China*

[8]*Materials Genome Institute, Shanghai University, Shanghai 200444, China*

**Abstract**
Mechanical THz vibrations in nanocrystals have recently been harnessed for quantum sensing and thermal management. The free boundaries of nanocrystals introduce new surface wave solutions, analogous to the seismic waves on Earth, yet the implications of these surface waves on nanocrystals have remained largely unexplored. Here, we use atomistic molecular dynamics simulations and experimental neutron spectroscopy to elucidate these THz-scale features in nanodiamond. Our key insight is that thermally induced Rayleigh surface phonons, which have a low group velocity and an amplitude that decays exponentially away from the surface, are responsible for the previously observed but unexplained linear scaling of the low-energy vibrational density of states in nanocrystals. Large thermal atomic displacements, relative to the nanoparticle radius, induce perpetual surface quakes, even at ambient conditions. Normalised to the radius, the surface displacement ratio in diamond nanocrystals exceeds that of the largest recorded earthquakes by a factor of $10^{5\pm1}$. We explicate how these dramatic Rayleigh waves coexist with other distinctive features including confined lattice phonons, soft surface modes, the acoustic gap, Love waves, and Lamb modes, thereby offering a complete framework for the vibrational dynamics of nanocrystals.

## Introduction

Travelling mechanical waves exhibit different behaviour at the surface of objects with free boundaries, as first shown by Rayleigh's famous mathematical solution in 1885.[1] In the field of seismology, this has tremendous consequences as earthquakes give rise to a mixture of body (bulk) and surface waves. The slower surface waves are primarily responsible for structural damage, due to their enhanced displacements. These surface features are also the basis for well-established technologies such as electronic devices that exploit surface acoustic waves (SAWs) to perform filtering at GHz frequencies, utilising wavelengths in the micrometre range.[2,3] Although Rayleigh SAWs have been intensively studied in the continuum limit,[1] their behaviour in ultra-small nanocrystals at nm-wavelengths, corresponding to THz frequencies, is far less understood. In this limit, atomistic and finite-size effects become important. Although theories widely predict Rayleigh waves in nanostructures,[4] direct experimental detection is challenging. With the trend towards ever-smaller technologies, the behaviour of short-wavelength (THz) vibrations at boundaries is of growing importance as these disproportionately influence the thermal properties of nanoarchitectures with large surface areas.[5,6] Thus, understanding surface vibrations has ramifications in the design of semiconductor devices,[5] thermoelectric materials,[7,8] topological phononic materials,[9] phononic diodes,[10] and waveguides.[11] The optomechanical coupling between lattice vibrations and electromagnetic radiation is particularly important in quantum computing,[12] telecommunications,[13] and sensors operating at frequencies in the GHz-THz range.[14]

Since nanocrystals (NCs) are a key component in emerging phononic,[15,16] sensing,[17] and catalytic systems,[18,19] a complete understanding of their vibrational properties is important. With reducing crystal size, surface vibrations become increasingly significant, occupying a larger fraction of the available modes. Some key NC features commonly observed include an excess of low energy "soft surface modes" and a reduction of phonon lifetimes due to scattering.[20,21] Reduced size also leads to phonon confinement effects which alter thermal properties when the phonon mean-free path becomes comparable to the size (*d*) of the object.[6] A distinct feature, known as an acoustic gap, has been predicted for wavelengths $\lambda > d$,[22] but is yet to be observed in practice. Meanwhile, deviations from the Debye law have been observed in the vibrational density of states (DOS) at low energy in many separate experiments, but not explained theoretically. In particular, the role of SAWs has not been elucidated in NCs. A unified picture of NC vibrational dynamics is therefore lacking.

*Corresponding Authors – D. L. Cortie: dcr@ansto.gov.au. and D. Yu: dyu@ansto.gov.au

Here, we study the case of nanodiamond (ND), a structurally simple NC particle with a monoelemental crystal lattice and highly harmonic atomic bonds. We show that simple, classical molecular dynamics (MD) simulations can be used to effectively and efficiently explore the unique vibrational modes of NCs and use inelastic neutron spectroscopy (INS) to validate the existence and nature of these modes in a real ND powder. In doing so, we identify the existence of Rayleigh SAWs in an NC for the first time and explain their impact on the vibrational DOS. These surface modes coexist with bulk lattice phonons and other unique ND modes, such as discrete, whole-particle resonances (Lamb modes), and likely have a significant impact on their thermodynamic and other properties. This study provides a map of nanoacoustic phenomena in NDs that may be useful in the design of ND- or other NC-based phononic, sensing, and catalytic systems.

## Results
### Unique Nanodiamond Lattice Vibrations from Molecular Dynamics
MD simulations show that characteristic surface features exist in nanoscale diamond that are distinct from the features in infinite bulk diamond (Fig. 1). Here, the MD velocity autocorrelation functions are used to calculate the vibrational DOS with no approximations other than the well-established classical carbon force-fields. Fig. 1 compares the vibrational modes of three simulated systems: 1) an idealised spherical ND (diameter, $d = 5.3$ nm), 2) the same ND particle with surface disorder from a semi-amorphous shell, and 3) a supercell of diamond with periodic boundary conditions modelling an infinite crystal. Several distinctions can be observed across energy regions (labelled *i-vii*), including different gaps in the DOS at low energy, highly resonant modes in the spherical ND, and broadened and softened (i.e., red-shifted) Van-Hove singularities in the NDs. The sub-sections below systematically discuss each of these unique features. The primary feature we wish to draw attention to is the large energy range (0–40 meV, labelled as *Region iii*) where a continuum of Rayleigh surface waves appears yielding a linear DOS which has not been explained in past work. To aid with the assignment of the features in the DOS, we also predict the INS dynamic structure factor, $S(Q, E)$, spherically averaged up to a distance of two Brillouin zones, as shown in Fig. 1b-d.

*Region i) Acoustic gap:* The largest allowable wavelength (lowest energy) phonon in a nanoparticle is determined by the length scale $2\pi/d$. Thus, an acoustic gap exists because of the forbidden region of $k$-space for wavelengths that exceed the particle diameter. We observe this confinement feature in the isolated NDs. We note that the existence of the gap in the periodic supercell is an artifact of the limited simulation box size and can be lifted by making the supercell larger (see Fig. S10). While the gap is an intrinsic feature of spatial confinement in an isolated NC, in realistic multiparticle solids, powders, or solutions, particle-particle interactions may also allow longer wavelengths to propagate, thereby "closing" the acoustic gap (as shown in Fig. S2).

*Region ii) Spheroidal Lamb mode resonances:* Resonant modes appear in the spherical ND, the strongest of which coincides with the band minima above the acoustic gap – notably lower in energy than the lowest energy mode for the supercell. In the $S(Q, E)$ map, these features appear as flat (non-dispersive) modes, characteristic of standing waves. The energy of the brightest feature matches well with the eigenfrequency of the fundamental Lamb mode (i.e., radial particle breathing) for an isotropic elastic sphere with size and elastic coefficients of the simulated ND.[23] This fundamental mode has been observed in simulations[24,25] and low-frequency Raman measurements[26,27] for NDs of similar sizes, and a summary is given in Table S2. According to Lamb's famous solutions,[23] weaker, higher-order modes are also present, each of which can be labelled by its angular momentum number (see Fig. S4). We observe that these modes are sensitive to structural disorder, as can be seen by the significant broadening of the fundamental and higher-order Lamb resonances in the surface-disordered core-shell (CS)-ND. This is consistent with the known sensitivity of breathing modes in NCs with respect to geometry.[24,28-30] We also find that interparticle interactions suppress these motions (see Fig. S2). These effects likely contribute to the Lamb modes being much less pronounced in experiment compared to what is calculated in Fig. 1a and Fig. 1b for ideal, isolated particles. We also note that the intensity of the Lamb resonance is sensitive to the MD simulation equilibration procedure (see the SI for a detailed discussion).

*Region iii) Rayleigh surface acoustic wave:* The energy region 0–40 meV contains the feature of primary interest in the current paper. Here, we observe a linear scaling of the DOS, $g(E) \sim E$, in the NDs, distinct from the quadratic, $g(E) \sim E^2$, scaling found in the bulk diamond simulation, which is expected from Debye's theory of solids.[31] The linear scaling has been observed in multiple NCs by neutron and Raman spectroscopy and simulations,[32,33] however, there has not yet been a definitive explanation of its origin. The $S(Q, E)$ map (Fig. 1b) provides additional insights: a single, linearly dispersive phonon mode is present, distinct from the known modes of infinite diamond (Fig. 1c). The linearly dispersive mode occurs at lower energy than the bulk acoustic modes due to its low group velocity, fitting the description of a Rayleigh wave; a primarily transverse SAW analogous to the slow surface component of an earthquake.[34] While "surface phonons" have been hypothesised as the origin of this excess DOS in other NCs, the proposal that these are Rayleigh waves is new, and no evidence has been given previously.

The unique characteristics of Rayleigh waves are highlighted by considering the dispersion, $\omega = c_R k$, where $c_R$ is the Rayleigh wave group velocity. It is well known that $c_R$ can be derived from the bulk wave velocities by solving:

$$\xi^6 - 8\xi^4 + \xi^2(24 - 16\eta^2) + 16(\eta^2 - 1) = 0, \quad (1)$$

where $\xi = c_R/c_t$ and $\eta = c_t/c_l$, with $c_t$ and $c_l$ corresponding to the bulk transverse and longitudinal waves, respectively.[1] We note that $\xi$ is uniquely given in terms of the Poisson ratio, $\nu$. For media with $0 < \nu < 0.5$, the Rayleigh SAW is slower than the bulk transverse wave, generally reduced by a factor of $0.87 < \xi < 0.96$.[35] In the MD simulations of the ND, $c_R$ is also reduced, giving a value close to the lower bounds of the classical expectation ($\xi \sim 0.73$) by Eq. 1 (see Fig. S5). We suggest that the additional softening of the SAW occurs due to the breakdown of the continuum approximation in the particle (finite-size effect). The excess of low-energy DOS is enhanced in the CS-ND, and we observe that the Rayleigh mode is significantly broadened in $S(Q, E)$, overlapping with the Lamb resonances. In a later section, we further dissect the Rayleigh wave contribution and show that it unambiguously originates from the surface and is responsible for the linear DOS scaling.

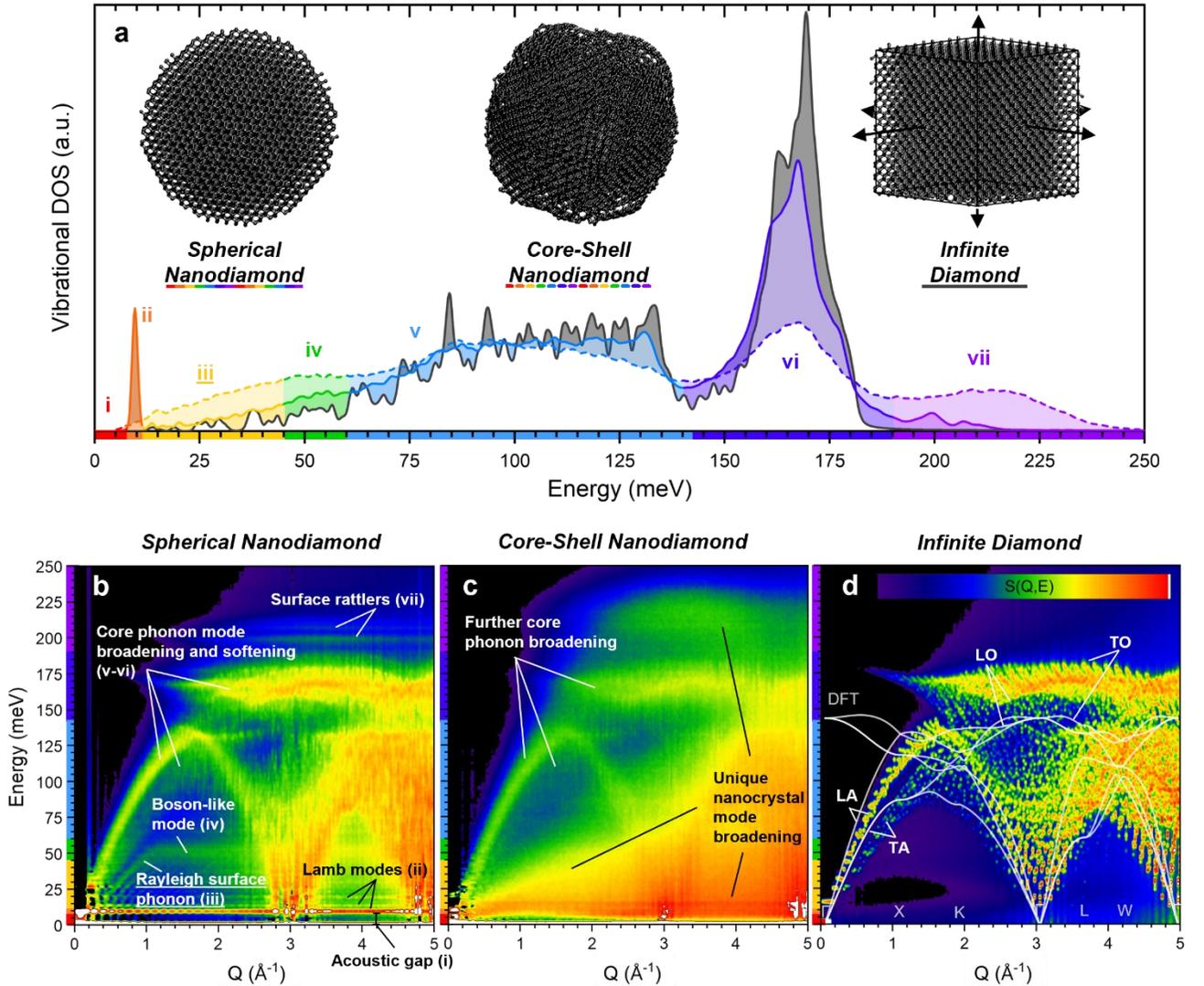

**Figure 1:** Molecular dynamics (MD) calculations reveal the unique vibrational properties of nanodiamond particles compared to bulk, infinite (periodic) diamond. a) The MD-calculated vibrational DOS for isolated diamond nanoparticles (spherical and core-shell, $d \approx 5.3$ nm) and a similarly sized periodic diamond supercell. The colours distinguish unique regions where there are differences between the supercell and particle simulations (*i-vii*). The neutron dynamic structure factors, $S(Q, E)$, calculated for the same systems are shown in b)-d). The log-scale intensity colour bar is given in (d), while the coloured energy axis corresponds to the energy regions in (a). In the $S(Q, E)$ maps, labels are provided for the unique modes of the spherical ND, their evolution in the core-shell particle, and the bulk modes from infinite diamond (clarified by [0 K] density functional theory lattice dynamics calculations). All MD simulations were performed at 300 K.

*Region iv) Boson-like peak:* A flat mode is present in $S(Q,E)$ for the NDs at ~50 meV, causing a bump in the DOS that is strongly enhanced in the CS-ND. This feature has the signatures of a "boson peak": a (broad) excess in the low-energy DOS over the contribution from normal modes (i.e., the Debye DOS), observed most commonly in disordered materials.[36] One long-standing view of the origin of the boson peak is from the competition between propagation and damping induced by the resonant coupling between transverse phonons and short-scale, quasi-localised vibrations with a transverse nature which are often induced by disorder.[37,38] In Fig. 1b, we see the localised mode responsible for the Van-Hove singularity intersecting with the Rayleigh SAW which has a strong transverse character. It is possible that the origin of this peak is related to the resonant coupling between the Rayleigh SAW and a high-order Lamb wave or disorder-induced surface mode with a transverse nature. We discuss this further in the SI. Future work is warranted to investigate this feature, especially given its potential influence on the thermodynamic properties of NCs.

*Region v) Modified bulk acoustic modes:* Broadening in the energy of the bulk diamond modes in the NDs compared to the periodic diamond is clear in the DOS and $S(Q,E)$. This is consistent with $E$-broadening due to phonon scattering induced by disorder and confinement in the NCs which leads to reduced phonon lifetimes. It is expected that the phonon scattering rate in nanostructures depends on phonon velocity and confinement size, with a scattering rate $\tau \sim \frac{d}{c}$ leading to a broadening of the linewidth $\Gamma \sim \frac{1}{\tau} \sim \frac{1}{d}$ occurring at the meV scale in NCs. We also see a subtle softening of the acoustic modes in the NDs. This is attributed to (radial) tensile forces in the NCs due to altered surface atom coordination which provides a sort of surface tension, somewhat similar to a liquid-vapour interface.[22]

*Region vi) Modified bulk optical modes:* Similar to the acoustic modes, the bulk diamond optical modes in the NDs are modified by significant lifetime broadening and mode softening. This has previously been observed via Raman spectroscopy.[39]

*Region vii) High-energy surface rattling modes:* Another unique feature of the NDs is the presence of high-energy optical modes. We label these as rattling surface modes. These arise from the existence of low-coordination ($sp^2$ or $sp$) surface carbons. It is expected that the surface ligands of NCs have vibrational or rotational modes at energies higher than the bulk phonons.[15] The modes are well-defined for the spherical ND which has only one variety of ligand-like surface carbon. However, the disordered surface of the CS-ND accommodates a wide range of surface atoms with different bonding and local environments and so the DOS is increased and broadened.

Some of the features discussed above have been identified in the past literature, albeit in isolated instances, and without a complete unified framework of the underlying physics. In particular, the observation that these phenomena overlap with SAWs in frequency and reciprocal space was not identified. In the remainder of the paper, we shall focus primarily on the new observation of Rayleigh SAWs in *Region iii)*.

**Isolating the Contribution of the Surface Acoustic Waves**
To show explicitly that the linearity in *Region iii)* arises from Rayleigh SAWs, we dissect the vibrational dynamics of the spherical ND by separating out the contributions from the core and surface, as shown in Fig. 2a (corresponding $S(Q,E)$ maps are shown in Fig. S6a-c). The core atoms of the ND strongly resemble the infinite diamond (dotted line), with quadratic Debye scaling of the low-energy DOS. In the outer shells, the DOS clearly transitions to a linear energy dependence. The corresponding $S(Q,E)$ maps confirm that the additional, linearly dispersive mode that we have attributed to a Rayleigh phonon is localised on the surface of the ND and provides the extra contribution to the DOS. Various other features (bulk diamond phonons, boson-like mode, and rattling modes) vary from core to surface, but none of these introduces a linear DOS.

A key characteristic of SAWs is that they only appear in the case of free boundary conditions. To demonstrate this, we imposed rigid (Dirichlet) boundary conditions on the ND, implying vanishing displacement across the ND surface (Fig. 2b). In this case, the excess low-energy DOS associated with the SAWs vanishes and the quadratic scaling of the DOS returns. Intuitively, the surface rattling modes are also frozen out under rigid boundary conditions, as is the fundamental Lamb resonance since the NC can no longer change volume. Fig. S6d-g shows that the ND with rigid boundary conditions closely resembles the periodic supercell $S(Q,E)$ in terms of the available phonon modes, while retaining the $E$-broadening of modes in the free ND. The rigid ND also has stiffened acoustic and optical modes compared to the free ND and infinite diamond. This has important implications for embedded NCs. Since the linear scaling of the DOS is also a characteristic feature of bulk liquids,[40,41] we also calculated the DOS of the NDs using instantaneous normal mode analysis (Fig. S8). Each calculation showed very few negative, unstable modes, ruling out the role of liquid-like dynamics as the origin of the linear DOS and supporting the SAW interpretation.

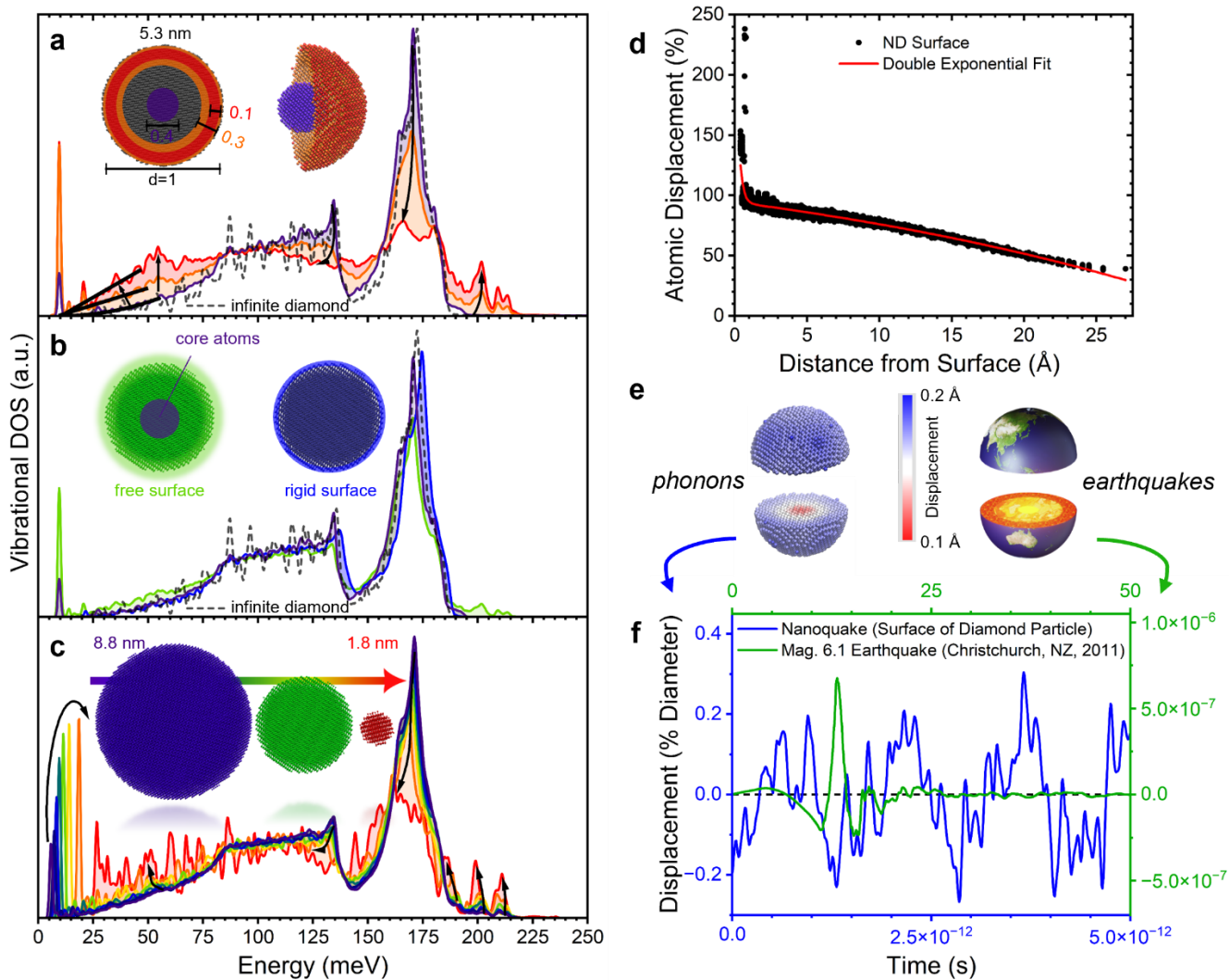

**Figure 2:** Dissection of the nanodiamond (ND) vibrational DOS reveals linear scaling originating from seismic-like vibrations on the free particle surface. a) ND vibrational DOS separated into contributions from the inner 40% (purple), outer 30% (orange), and outer 10% (red) diameter. b) Total (green) and core-atom (purple) DOS for the free particle compared with the total DOS of the same ND under Dirichlet boundary conditions (blue). Infinite diamond is given for reference (dotted line) in (a) and (b). c) DOS size comparison for spherical NDs ranging from ∼1.8–8.8 nm in diameter (corresponding periodic supercell DOS are provided in Fig. S10) (exact sizes are given in Table S1). d) Atomic displacements in the 5.3 nm spherical ND show (bi)exponentially decaying amplitude from the surface of the particle. e) The SAW phonons in the ND are compared with earthquake waves. f) The relative surface displacement (with respect to object diameter) of the Rayleigh phonons (ND) and seismic waves (Earth) are compared.

As the ratio of the number of surface to core atoms scales as $\sim 1/d$, we also simulated the ND particle size-dependence of the total DOS (Fig. 2c) ($S(Q,E)$ maps shown in Fig. S6h-j). The evolution of modes from large to small particles bears a strong resemblance to the evolution of core to surface atoms. In addition to enhancing the linear contribution, the size-dependence shows clear softening of optical modes, attributed to increasing tensile stresses in the smaller particles. Multiple confinement effects are also highlighted in the size-evolution of the DOS: the acoustic gap scales as $\sim 1/d$, as does the resonant breathing mode and linewidth broadening. The acoustic gap and Lamb resonance energies match well with elasticity theory (Fig. S9).[30]

When mapping the displacement of atoms as a function of their distance from the ND surface (Fig. 2d), we observe exponentially decaying motion – another characteristic feature of Rayleigh waves.[1,34] More specifically, the bi-exponential behaviour is consistent with two SAW phonon mode-polarisations, an elliptical (Rayleigh) and an in-plane (Love) contribution. The behaviour of SAWs in NCs can be compared to the more familiar (seismic) motions of macroscopic objects (Fig. 2e,f).

The data in Fig. 2 and Fig. S6 demonstrate that the linear DOS originates from the Rayleigh SAWs and highlight its 2D nature, being confined to the surface of the ND particle. The low-energy DOS of isolated gold particles was also recently studied and demonstrated to transition from quadratic dependence for large particles to linear dependence for small particles.[32]

In this work, an *a priori* link was proposed between the linear DOS and the dimensionality of the NC surface, suggesting that since the Debye DOS scales as $g(E) \sim E^{n-1}$, where $n$ is the dimensionality of the system, the linear DOS likely originates from the "2D" NC surface. The observation of 2D, THz-scale Rayleigh SAWs in this work supports this hypothesis and provides a clear mechanism.

So far as we can tell, there has not been any formalism created for the contribution of Rayleigh SAWs to the phonon DOS. However, there is a simple heuristic argument that suggests confined SAWs lead to an effective 2D DOS. Let us consider an elastic medium with a surface extending along the $x, y$ directions, neglecting possible effects arising from the curvature of such a surface, so that Cartesian coordinates can be locally adopted. This is appropriate for THz acoustic phonons which have a short wavelength ($\ll$ 1 nm). Rayleigh phonons are excitations that propagate near the surface without penetrating the body. In other words, their wave-vector along the $z$-direction, perpendicular to the surface, is imaginary, implying their evanescent decay inside the elastic medium. On the other hand, the wave-vectors along the $x, y$ directions are real, and the Rayleigh phonons behave as standard normal modes along the surface. The corresponding DOS can be derived by counting the modes along the surface directions along which the Rayleigh phonons propagate:

$$g(\omega)d\omega = N_A\, dk_x dk_y = 2\pi N_A k\, dk, \qquad (2)$$

where $N_A$ is the number of states per unit area in $k$-space, as in the Debye model. Using the linear dispersion relation of Rayleigh phonons, and the relation $E = \hbar\omega$, the DOS is then given by:

$$g(E) \sim \frac{E}{(c_R)^2}. \qquad (3)$$

The linear, 2D-DOS is thus proposed to be a unique feature of NCs hosting specific phononic surface states with radial exponential decay.

**Experimental Confirmation of the Linear DOS from Rayleigh Waves**

To experimentally confirm the existence of SAWs in NCs, we measured and analysed the DOS and $S(Q, E)$ for ND and microdiamond (µD) powders utilising time-of-flight and triple-axis INS experiments. The INS-derived DOS for the ND and µD powders, as well as the MD-derived DOS for the 5.3 nm ND for reference, are shown in Fig. 3a. There is good agreement between most of the features across the (*i-vii*) energy regions of the MD DOS and the INS DOS when energy scaling is applied to offset limitations in the MD force-constants and smoothing is applied to account for the instrument energy resolution. TEM micrographs of the µD and ND powders are shown in Fig. 3b and 3c, respectively.

The low-energy scaling of the DOS is shown in Fig. 3d. The data in this region are fitted with a second-order polynomial and Gaussian function. The µD sample shows a clear quadratic energy dependence, while the ND sample shows an unambiguous linear scaling between ~0-10 meV, with the linear component dominating up to 40 meV, consistent with the prediction of Rayleigh SAWs. These features are highlighted in Fig. 3e, where the DOS has been scaled by $1/E$.

While the individual vibrational modes in the experimental $S(Q, E)$ maps are more difficult to isolate compared to the simulated ones, clear differences between the µD (Fig. 3f) and ND (Fig. 3g) are observed, including signal that is reminiscent of quasielastic neutron scattering (QENS) in the NDs. We believe that this additional signal in the ND is primarily from the surface modes that give rise to the linear DOS. A clear resemblance can be seen between this signal and the simulated surface modes that are broadened in the CS-ND (Fig. 1c), which include the Rayleigh mode. While this signal also resembles the calculated inter-particle and diffusional dynamics shown Fig. S2b (actual QENS), the contribution to the DOS does not correlate. We do not rule out some contributions from these dynamics but suggest that they are minor, especially at energies above a few meV.

Aside from the clear linearity in the DOS, which indicates the presence of Rayleigh phonons, there are other qualitative differences observed between the ND and µD powders, matching the predictions of the MD. A broad peak is observed at ~31 meV, close to the feature in the MD DOS (boson-like peak, *Region iv*). The deviation of the DOS at ~31 meV from the linear scaling in the ND sample is highlighted in Fig. 3e. Excess modes (surface rattlers) are also evident beyond the bulk phonon cut-off. While the fundamental Lamb (breathing) mode is not clearly visible in the INS DOS, we suspect that this is due to the particle morphology (core-shell), inter-particle coupling, and size inhomogeneity causing the mode to be suppressed and broadened. Stiffening of the optical modes is observed in the ND INS DOS compared to the µD INS DOS, another feature that we have observed originating from interparticle interactions in ND-cluster simulations (Fig. S2). Using triple-axis spectroscopy, we also observe significant broadening of the TA phonons in the ND compared to the µD (note that the µD linewidth is instrument-resolution-limited), consistent with our observations in the simulations which are suggestive of reduced phonon lifetimes due to confinement and disorder (Fig. 3h). Additionally, we observe softening of the TA phonons in the ND in line with the observations of the single-particle ND MD simulation from radial tensile stress due to the particle surface. The triple-axis scans also reveal a mode in the ND at ~19 meV which is lower in energy than the TA phonon, suggesting that it is a unique NC mode. Correlating the ($Q,E$) location of this mode (Fig. SI2), we assign this as a high-order Lamb mode.

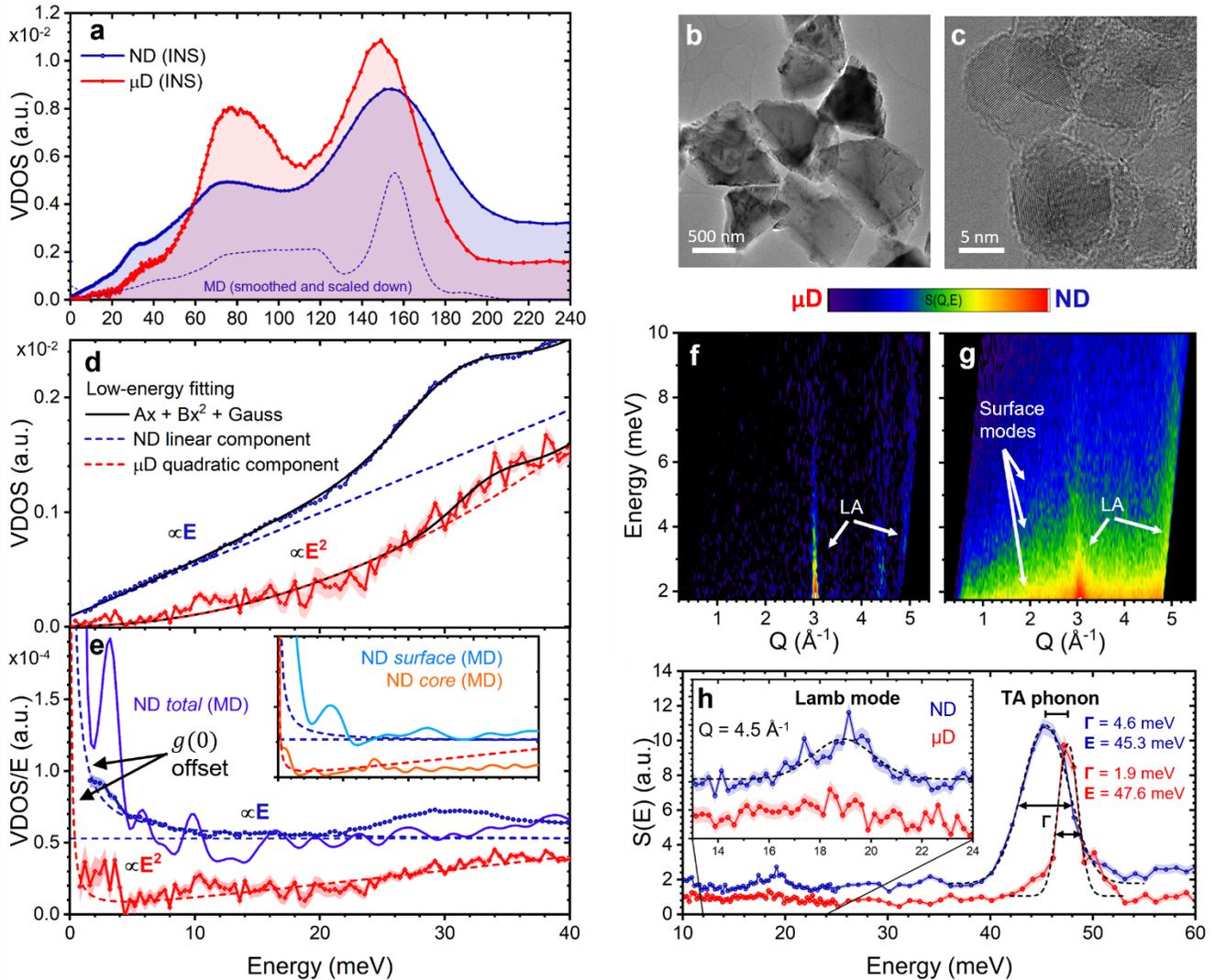

**Figure 3:** Inelastic neutron scattering experiments confirm unique vibrational modes in nanodiamond compared with microdiamond including the linearly scaling DOS due to Rayleigh SAWs predicted by atomistic simulations. a) The neutron vibrational DOS of the µD and ND powders. TEM micrographs are shown for the b) µD and c) ND powders. The semi-amorphous shell and crystalline core of the NDs are clearly observed. d) The low-energy region of the DOS highlights the linear vs quadratic scaling of the ND and µD samples as well as the boson-like peak DOS at ~31 meV. These features are further highlighted by e) the same DOS scaled by $1/E$. The MD-calculated DOS for the spherical ND shows the same behaviour as the real powder, and the inset shows the quadratic scaling of the core atoms of the MD particle and the linear scaling of the surface atoms – the tick values on both axes are the same as the main plot. The dashed lines correspond to the $E$ and $E^2$ fits in (d) (the horizontal line is the same slope through $g(0) = 0$). The neutron $S(Q, E)$ maps for the f) µD and g) ND samples at low energy near the Bragg peak ($Q = 3.05$ Å$^{-1}$) show the origin of the excess ND DOS. h) Triple-axis energy scans at $Q = 4.5$ Å$^{-1}$ for µD and ND reveal TA phonon energy (E) softening and linewidth (Γ) broadening in the ND and evidence of a high-order Lamb mode (inset). Instrumental error in the line plots is given by the shaded area above and below the data points. All INS data were taken at 583 K.

## Conclusions

Our data show that, in addition to changes to core phonons, the surfaces of NDs undergo persistent violent thermal nanoquakes. Compared to macroscale earthquakes, these nanoquakes (i.e., Rayleigh phonons or SAWs), are ultra-fast (10$^{-12}$ vs 10$^1$ second scale) and ultra-large (relative to object diameter; 10$^{-1}$ vs 10$^{-7}$ % diameter) (Fig. 2f). Rayleigh phonons, which we have demonstrated as the origin of the linearly scaling DOS in NDs, are expected to have a variety of important implications. For example, the existence of Rayleigh waves in other nanostructures has been shown to significantly alter thermal conductivity through altered phonon scattering and surface localisation.[4] NC surfaces also mediate mechanical vibrations between the NCs and their local environments. The existence of large amplitude, dispersive surface modes will most likely alter this interaction, as suggested by [4]. A local chemical environment that allows for the existence of such SAWs (perhaps in liquids or other soft materials) may have either improved thermal energy transfer with NCs (due to the excess of available modes) or degraded thermal transfer (due to enhanced phonon scattering and localisation). These considerations are important for thermal management and sensing applications such as thermometry. Changes in the amplitude and frequency of NC surface modes in different chemical environments may also provide crucial tools to study the bonding interactions of NCs with their local environments. It is also expected that these proportionally large surface displacements will have a significant impact on catalysis, and NC surface vibrational effects should be given careful attention, as is well-outlined in [42]. More generally, unique NC dynamics will alter the thermodynamic properties (such as heat capacity, vibrational entropy, and thermal expansion) of NCs. Such properties are important for the applications mentioned and beyond.[43] Further studies on the precise influence of these modes on NC heat capacity will likely be of particularly high value. Aside from the suspected general implications of our results for NCs, the ND particles in our work also have specific technological relevance to diamond-based thermometry,[44,45] quantum computing,[46] and thermal engineering.[8,47,48] We hope that the comprehensive framework of ND vibrational dynamics presented in our work will encourage further investigations into the roles that nanoacoustic modes play in various applications.

## Methods
### Classical Molecular Dynamics
Classical molecular dynamics (MD) simulations were run using the LAMMPS molecular dynamics simulator.[49] Each production simulation was run using a canonical (NVT) ensemble with temperature fixed at 300 K or 583 K. In all ensembles, pressure and temperature were controlled using the Nose-Hoover method.[50,51] A timestep of 1 fs was used for all runs. For production, frames were dumped at a frequency of 1/3 frames/fs (every 3$^{rd}$ frame) to be processed for the vibrational density of states and coherent dynamic structure factor.

For the ND particles, a spherical particle was created in a large vacuum box (~140 nm) and energy-minimised. Atom velocities were adjusted so that the net linear and angular momentum of the particles was zero. The isolated particles were simulated with a 10 ps NVT run (2 ps equilibration, 8 ps for production). As discussed in the SI, this procedure leads to a large population of Lamb modes, which are less populated if more aggressive minimisation and equilibration procedures are undertaken, though the other vibrational modes are unaffected.

Another particle with a graphitic/amorphous shell (core-shell, CS-ND) was created by annealing the 5.3 nm diamond particle at 850 K for 2 ns and then quenching to 300 K for 1 ns using NVT ensembles.

Spherical diamond particles of 8 sizes, ranging from 1.8-8.8 nm in diameter were investigated and supercells of the corresponding number of atoms (4x4x4 to 20x20x20 supercells) with periodic boundary conditions were studied for reference (see Table S1 for details). To achieve stable pressure in the periodic "infinite" diamond, an additional initial isothermal-isobaric ensemble (NPT) run was simulated with pressure fixed to 1 atm. The AIREBO (without Lennard-Jones or torsion contributions) potential was used for all carbon-carbon interactions.[52]

Finally, clusters of the smallest ND (1.2 nm) were simulated to demonstrate the role of inter-particle interactions. Particles (2, 3, 5, 10, and 15) were placed near each other (<10 Å) and the simulation run for enough time for the particles to interact and form a stable cluster (~20 ns).

To impose Dirichlet boundary conditions on the nanoparticles, the same simulations were run with the surface atoms fixed with force equal to zero. The full list of systems investigated with MD is presented in Table S1. Visual representations of the MD simulations were created using Visual Molecular Dynamics (VMD).[53]

While classical simulations inherently lack the ability to include quantum effects, undertaking simulations on such large particles using alternative *ab initio* methods is currently not feasible. One consequence of this is that the population of vibrational modes is not correct (see Fig. S2), which will have some quantitative effect on the linewidth broadening, particularly at low temperatures where Bose-Einstein population effects are important. Nonetheless, we hope that this work highlights the power of using classical MD for studying THz dynamics in nm-sized systems at elevated temperatures and that with the advancement of machine-learning atomic potentials, semi-classical simulations with improved force constants will offer even more accurate results.

### Phonon Density of States and Dynamic Structure Factor Calculations (Simulation and Neutron Scattering Theory)
The resulting trajectory files from the MD simulations were used to calculate the dynamic quantities presented throughout this work. These calculations were performed using the Molecular Dynamics Analysis of Neutron Scattering Experiments (MDANSE) program.[54] Since MDANSE simulates neutron scattering data, it is logical to first provide the theory and background in terms of neutron scattering (with neutron experimental details provided later) and then detail how MDANSE computes the corresponding data from the MD trajectories.

The density of states (DOS) data are derived utilising the velocity autocorrelation function (VACF), i.e., by integrating, over time, all the positional and velocity information of the system. The VACF for an atom in an atomic system is defined as

$$C_{vv;\alpha\alpha}(t) \doteq \frac{1}{3}\langle v_\alpha(t_0) \cdot v_\alpha(t_0+t)\rangle_{t_0}, \tag{4}$$

where $v_\alpha(t)$ is the velocity of atom $\alpha$ as a function of time. For an isotropic system, the DOS (as a function of angular frequency) can be given as

$$g(\omega) = \sum_\alpha b_{\alpha,inc}^2 \tilde{C}_{vv;\alpha\alpha}(\omega), \tag{5}$$

where $b_{\alpha,inc}$ is the incoherent (i.e., single-particle motion) neutron scattering length for atom $\alpha$, and

$$\tilde{C}_{vv;\alpha\alpha}(\omega) = \frac{1}{2\pi}\int_{-\infty}^{+\infty} e^{-i\omega t} C_{vv;\alpha\alpha}(t)\, dt. \tag{6}$$

MDANSE computes the discrete DOS (using the un-normalised VACF so that $g(0)$ approximates the diffusion constant) by:

$$DOS(n \cdot \Delta v) \doteq \sum_\alpha \omega_\alpha \tilde{C}_{vv;\alpha\alpha}(n \cdot \Delta v), n = 0 \ldots N_t - 1, \tag{7}$$

where $N_t$ is the total number of time steps and the frequency step is $\Delta v = 1/(2N_t\Delta t)$. This discrete DOS is then smoothed by applying a Gaussian window (in $t$), producing the simulated $g(\omega)$.

$S(Q,\omega)$, which contains information about the structure and dynamics of the scattering system, is given by the Fourier transform of the intermediate scattering function, $F(Q,t)$, as follows.

$$S(Q,\omega) = \frac{1}{2\pi}\int_{-\infty}^{+\infty} e^{-i\omega t} F(Q,t) dt, \tag{8}$$

$$F(Q,t) = \sum_{\alpha,\beta} \Gamma_{\alpha\beta} \langle e^{-iQ \cdot \hat{R}_\alpha(0)} e^{-iQ \cdot \hat{R}_\beta(t)}\rangle, \tag{9}$$

where

$$\Gamma_{\alpha\beta} = \frac{1}{N}\left[\overline{b_\alpha b_\beta} + \delta_{\alpha\beta}\left(\overline{b_\alpha^2} - \overline{b_\alpha}^2\right)\right]. \tag{10}$$

Here, $Q = |Q|$, where $Q = k_i - k_f$ is the scattering vector and $\omega = (E_0 - E)/\hbar$ is the corresponding scattering event energy transfer. The operators $\hat{R}_\alpha(t)$ in Eq. 9 are the position operators of the nuclei in the sample. The brackets $\langle \ldots \rangle$ denote a quantum thermal average and the time dependence of the position operators is defined by the Heisenberg picture. In classical MD, these are replaced with the classical position, and an approximate detailed balance factor is applied. The symbol $\overline{\ldots}$ denotes an average over isotopes and relative spin orientations of the neutron and the nucleus. In a neutron scattering experiment, both coherent (collective atomic motion) and incoherent (individual atomic motion) components of $F(Q,t)$ (and hence, $S(Q,\omega)$) will be present, according to the atomic scattering lengths $b_{\alpha,coh} \doteq \overline{b}_\alpha$ and $b_{\alpha,inc} \doteq \sqrt{\overline{b_\alpha^2} - \overline{b_\alpha}^2}$, respectively. MDANSE computes the coherent dynamic structure factor by first computing the coherent intermediate structure factor on a rectangular grid of equidistantly spaced points along the time- and the $Q$-axis, respectively:

$$F_{coh}(Q_m, k \cdot \Delta t) \doteq \sum_{I=1, J \geq I}^{N_s} \sqrt{n_I n_J \omega_I \omega_J}\, \overline{\langle \rho_I(-Q,0)\rho_J(Q, k \cdot \Delta t)\rangle^Q}, \tag{11}$$

$$k = 0 \ldots N_t - 1, m = 0 \ldots N_Q - 1,$$

where $N_t$ is the number of time steps in the coordinate time series, $N_Q$ is a user-defined number of Q-shells, $N_s$ is the number of selected species, $n_I$ the number of atoms of species $I$, $\omega_I$ the weight for species $I$, $\rho_I(Q, k \cdot \Delta t)$ is the Fourier transformed particle density for species $I$, and the symbol $\overline{...}^Q$ denotes an average over the Q-vectors. We used Q-vectors for a spherical lattice (Q shells) from $Q = 0$–$5$ Å$^{-1}$, in steps of 0.025 Å$^{-1}$, with a maximum of 50 $hkl$ vectors per shell. The Fourier transformation of $F_{coh}$ is then calculated (as per Eq. 11) to produce $S_{coh}$. As per the DOS calculation, a Gaussian smoothing procedure is implemented on the discrete data. An "instrument resolution" of 1 THz was chosen for $g(\omega)$ and $S(Q, \omega)$. We report $S$ and $g$ as a function of energy (transfer) in the form $E = \hbar\omega_i - \hbar\omega_f$.

**Instantaneous Normal Mode Analysis**
DOS calculations were also computed from the same trajectory outputs using the instantaneous normal mode (INM) approach (custom code). The INMs are determined through diagonalisation of the dynamical Hessian matrix at an instant of time. For each configuration, the Hessian matrix is a 3N × 3N matrix that assesses the second derivatives of the potential energy. Its elements are constructed as follows:

$$H_{i\mu,j\nu}(\boldsymbol{R}) = \frac{1}{\sqrt{m_i m_j}} \frac{\partial^2 V}{\partial r_{i,\mu} \partial r_{j,\nu}}, \tag{12}$$

where $i, j = 1, ..., N, \mu, \nu = x, y, z$. $\boldsymbol{R} \equiv \boldsymbol{r}_1, ..., \boldsymbol{r}_N$ represents each configuration and $\boldsymbol{r}_i$ is the position of the $i$th atom. $V$ is the potential energy and $r_{i,\mu}$ represents the $\mu$-coordinate of the $i$th atom. The instantaneous normal mode frequencies $\omega_i$ are the square roots of the eigenvalues of the dynamical matrix. The negative curvature of the potential energy landscape would lead to negative eigenvalues corresponding to purely imaginary frequencies, which are labelled as unstable INMs.[55] The INM spectrum is obtained by:

$$<\rho(\omega)> = <\frac{1}{3N}\sum_i^{3N} \delta(\omega_i - \omega)>. \tag{13}$$

For each structure, we performed instantaneous normal mode analysis for 100 different configurations generated at 0.1 ps intervals during the last 10 ps of the full simulation. The INM spectrum was averaged over the analysed ND configurations.

**Samples for Neutron Experiments**
Two commercially obtained samples were used for inelastic neutron scattering measurements: detonation synthesised nanodiamond powder with an average particle diameter of ~5 nm was obtained from US Research Nanomaterials, Inc., and synthetic monocrystalline diamond powder with particle size ~1 μm was obtained from Sigma-Aldrich (Merck). Additional TEM micrographs and X-ray and Neutron diffraction patterns are shown in Fig. SI 8. Approximately 5 g of ND sample and 5 g of μD sample were loaded into aluminium cans. Since NDs are highly hygroscopic,[56] we heat-treated the ND samples at 300 °C for ~20 hr in an air furnace and sealed them in their sample cans while still warm in the furnace. In previous experiments without heat-treating the NDs, a clear peak in the DOS from H$_2$O was observed, even at high temperatures while the sample can was sealed. The μD sample was heat-treated the same way for consistency.

**Neutron Spectroscopy**
Inelastic scattering measurements were carried out on two neutron spectrometers – Pelican and Taipan at the Australian Nuclear Science and Technology Organisation (ANSTO). All $g(E)$ and $S(Q, E)$ data were measured on the time-of-flight cold neutron spectrometer – Pelican.[57] An incident neutron energy of 3.7 meV (wavelength of 4.69 Å) was used. This offers a high flux and a good energy resolution of 0.13 meV at the elastic line. The DOS was obtained from the energy gain side. An incident energy of 14.9 meV, corresponding to the second-order reflection of the monochromators (half of 4.69 Å), was also used to directly visualise the acoustic waves at high $Q$ at the energy gain side. Measurements were taken at 583 K to maximise the scattering signal on the neutron energy gain side without having to worry about the softening of the aluminium can.

The raw neutron scattering data from Pelican was processed using the Large Array Manipulation Program (LAMP).[58] The basic physical principles of the measurement of $S(Q, E)$ are given in Eq. 8 and the associated text (precise details of how LAMPS processes the raw data into $S(Q, E)$ are provided in ref [59]). Once $S(Q, E)$ has been reduced from the data, the phonon DOS can be calculated according to

$$g(\omega) = C \int \frac{\omega}{Q^2} S(Q, \omega)(1 - e^{-\hbar\omega/k_B T}) dQ, \tag{14}$$

where $C$ is a factor containing the atomic mass and Debye-Waller factor, $e^{-2W}$, which is taken as unity for all samples here. The integration over $Q$ covers the experiment-accessible range of momentum transfer. Experimentally, the DOS derived from (14) corresponds to the generalised DOS which is the weighted value over the neutron scattering cross-section for individual elements contained in the sample.

Taipan, the thermal triple-axis spectrometer at ANSTO,[60] was used to perform constant-$Q$ energy scans on the samples. Taipan has access to neutrons with energy up to 200 meV by employing both a PG monochromator ($5 < E_i < 70$ meV) and a Cu monochromator ($25 < E_i < 200$ meV). Taipan was aligned with o-40'-40'-o collimation to ensure sufficient $Q$-resolution with a fixed final energy $E_f = 14.87$ meV giving an energy resolution of 0.9 meV (FWHM) at the elastic line for the PG monochromator. The samples were measured in a cryostat at 300 K.

## Funding
This project was primarily funded as a part of the Australian Research Council (ARC) Discovery Project (DP) DP210101436. Neutron beam and scientific computing times were awarded at ANSTO under proposals P14140 and P15658. CS is supported by the postgraduate research award (PGRA) provided by the Australian Institute of Nuclear Science and Engineering (AINSE), the Australian Government Research Training Program (AG-RTP), and the aforementioned DP. The high-performance computing aspect of this work was supported by the Adaptor grants scheme, with computational resources provided by NCI Australia, an NCRIS-enabled capability supported by the Australian Government, and by the Pawsey Supercomputing Research Centre's Setonix Supercomputer (https://doi.org/10.48569/18sb-8s43), with funding from the Australian Government and the Government of Western Australia.

## Author Contributions
D.C., D.Y., and C.S. conceived the original idea. C.S., D.C., and P.G. performed the MD simulations. C.S., D.Y., and K.R. performed the neutron scattering experiments. C.S., D.C., D.Y., and M.B. performed the formal data analysis. K.P. and P.G. performed the DFT simulations. S.J. and X.F. performed the INM analysis. A.B. and C.S. took the TEM micrographs. C.S., D.C., R. L, D.Y., and M.B. wrote and revised the manuscript draft. All authors contributed to the scientific discussion and revision of the manuscript.

## Acknowledgements
The authors would like to thank Dr John Osborn from ANSTO for his helpful comments regarding heat-treating the nanodiamond samples.

## Data Availability
LAMMPS (MD) input files and neutron scattering data can be found in the following repository:
https://url.au.m.mimecastprotect.com/s/nbzkCoV122HAoQxxh1fKtpKWZb?domain=github.com
Further data supporting the findings of this study are available from the corresponding authors upon request.

## Ethical Declarations
*Competing Interests*
The authors declare no competing financial interests.

*Supplementary Information for*

# Thermal Nanoquakes: Terahertz Frequency Surface Rayleigh Waves in Diamond Nanocrystals

C. Stamper,[1] M. Baggioli,[2,3] P. Galaviz,[4] R. A. Lewis,[5] K. C. Rule,[4,5] A. Bake,[1,6] K. A. Portwin,[1] S. Jin,[2,3] X. Fan,[7,8] D. Yu,[4*] and D. L. Cortie[4,5*]

[1]*Institute for Superconducting and Electronic Materials, University of Wollongong, Wollongong, NSW 2500, Australia.*

[2]*School of Physics and Astronomy, Shanghai Jiao Tong University, Shanghai 200240, China*

[3]*Wilczek Quantum Center, Shanghai Jiao Tong University, Shanghai 200240, China*

[4]*Australian Nuclear Science and Technology Organisation, Lucas Heights, NSW 2234, Australia.*

[5]*School of Physics, University of Wollongong, Wollongong, NSW 2500, Australia.*

[6]*Institute for Glycomics, Griffith University, Gold Coast, QLD 4215, Australia.*

[7]*Shanghai National Center for Applied Mathematics, Shanghai Jiao Tong University, Shanghai 200240, China*

[8]*Materials Genome Institute, Shanghai University, Shanghai 200444, China*

*\*Corresponding Authors: D.C.; dcr@ansto.gov.au. D.Y.; dyu@ansto.gov.au*

## Discussion

**Further Discussion on the MD and INS Lamb Mode Intensity**

As mentioned in the main text, we have found the intensity of the Lamb resonance in the spherical ND to be dependent on the simulation setup. The spherical ND simulations in this work are initialised with an energy minimisation that forces energy tolerance to $1 \times 10^{-4}$ and then run as NVT with a target temperature (e.g., 300 K). In this case, it takes about 2 ps for the temperature to equilibrate. We found that more aggressive energy minimisation and longer and more rigorous equilibration procedures significantly alter the Lamb resonance intensity. Fig. S1a shows the effect on the VDOS of altering the simulation run time with the same energy minimisation. Fig. S1b shows the effect on the VDOS of implementing a more aggressive minimisation procedure and combining that with a rigorous equilibration procedure. The differing intensity of the mode with different set-ups could suggest interesting non-equilibrium behaviour of the Lamb modes, although it is unclear whether the aggressive equilibration/minimisation artificially damps the Lamb resonances. The large dependence of the resonance intensity on the simulation set-up likely explains why some MD simulations of nanoparticles do not clearly observe these resonances if aggressive minimisation and equilibration parameters are utilised.[1,2] While the intensity of the Lamb mode may be smaller at complete equilibrium, the apparent "quasi-equilibrium" conditions in the simulations of this work still highlight the unique ND Lamb modes – which are clearly observed in experiments (see Table S2) – without jeopardising the observation of other NC modes, as evidenced by Fig. S1.

We also note that while the core-shell particle simulations are necessarily run with a longer equilibrium process due to the prior annealing steps, the Lamb resonances are still clearly observable in the $S(Q,E)$ map (see Fig. 1c in the main text and Fig S7). From additional simulations, we see no change when artificially offsetting this equilibrium to mimic the conditions of the untreated, spherical NDs. The seemingly non-equilibrium behaviour of the diamond Lamb modes observed in the MD raises interesting questions, e.g., about how thermally populated Lamb resonances are in real NDs at thermal equilibrium. This has potential implications for their observation in experiments that require the modes to be populated, such as the neutron energy gain data used to derive the VDOS in this work. Lamb modes in gold nanoparticles have been clearly observed in both Stokes (photon energy loss) and anti-Stokes (photon energy gain) Raman measurements.[3] The nature of these Lamb resonances in terms of their non-equilibrium behaviour should be investigated in future work.

**Further Discussion on the Boson-Like Mode**

In the main text, we discuss the presence of a prominent feature in the neutron nanodiamond (ND) density-of-states (DOS) at ~ 31 meV) (Fig. 3a,d,e, also Fig. S12d) that looks similar to that of the boson-like mode observed in the simulated nanodiamond particles and occurs at a similar energy. It is likely that the experimental signal has the same origins as the MD feature. However, the precise origin of this mode is unclear. In the main text, we suggest that it is possible that the origin of this mode is the resonant coupling between the Rayleigh SAW and a high-order Lamb wave or other surface mode with a transverse nature. This is one view that arises from the analysis of the $S(Q,E)$ map of the spherical ND (Fig. 1b). A long-standing view of the origin of the boson peak is the competition between propagation and damping induced by the resonant coupling between transverse phonons and short-scale, quasi-localised vibrations with a transverse nature, often induced by disorder. In references [4] and [5], Raman-active modes at the same energy (230-260 cm$^{-1}$, 28-32 meV) have been observed in radiation-damaged diamond after annealing. The authors attribute these modes to vibrations from "intermediate" carbon phases consisting of both *sp*$^3$ and *sp*$^2$ bonded carbon. These defects in the diamond structure are consistent with the spherical ND surface, which contains *sp*$^3$ and *sp*$^2$ bonded carbons, and especially with the annealed core-shell diamond particle which contains many *sp*$^3$ and *sp*$^2$ bonded carbons. Recall that the CS ND (and INS) DOS has an even stronger boson-like mode correlating with a greater portion of carbons in an intermediate phase. Therefore, it seems likely that the excess signal comes from these intermediate phase carbon atoms – through the resonant coupling of SAWs and/or Lamb resonances, or otherwise. When evaluating the DOS of the μD, a (smaller) excess of signal can also be observed centred at the same energy as the ND boson-like mode. It is possible that a small fraction of intermediate phase carbon is present in the μD sample as contamination (which is very difficult to detect via diffraction), and so we suspect that this signal also arises from this intermediate carbon in the μD DOS. In our TEM inspection of the μD sample, we can see nano-carbon inclusions that have similarly amorphous edges. If the peak is related to Rayleigh or Lamb resonances, it should be noted that neither necessitates spherical geometry, although the geometry will affect the eigenfrequencies of some resonances. Future experimental studies are required to elucidate these points.

# Figures

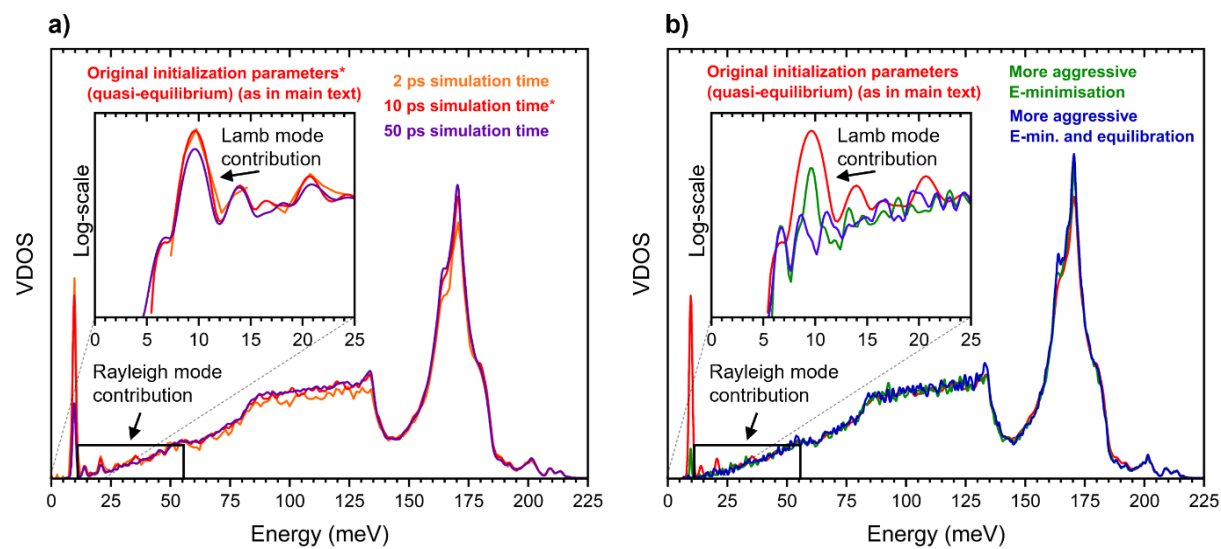

**Figure S1:** Comparison of the VDOS of a spherical nanodiamond (ND5) from MD simulations with different initialisation parameters. a) VDOS of ND5 simulated with the same minimisation (energy tolerance $1 \times 10^{-4}$) but different total simulation times: 1/5 of the original simulation (2 ps, orange) 5 times the original simulation, (50 ps, purple), and the original simulation (10 ps, red). Frames after the first 2 ps are used for the 10 and 50 ps simulations and all frames are used for the 2 ps simulation. b) VDOS of the same particle with different minimisation and equilibration procedures: the original conditions as in the main text and (a) (red), with a more aggressive energy minimisation (force tolerance $1 \times 10^{-8}$) (green), and with the more aggressive minimisation and a rigorous equilibration procedure; 50 ps of NPT heating with, followed by 1 ps of NVT holding at temperature with 0.2 drag (blue). The inset shows the changes in the fundamental Lamb mode with a log-scale y-axis. The Lamb resonance intensity decreases with longer simulation time (equilibrium) and with more aggressive minimisation and equilibration parameters. The differing intensity of the mode with different set-ups could suggest non-equilibrium behaviour, although it is unclear whether the aggressive equilibration/minimisation artificially damps the mode resonance.

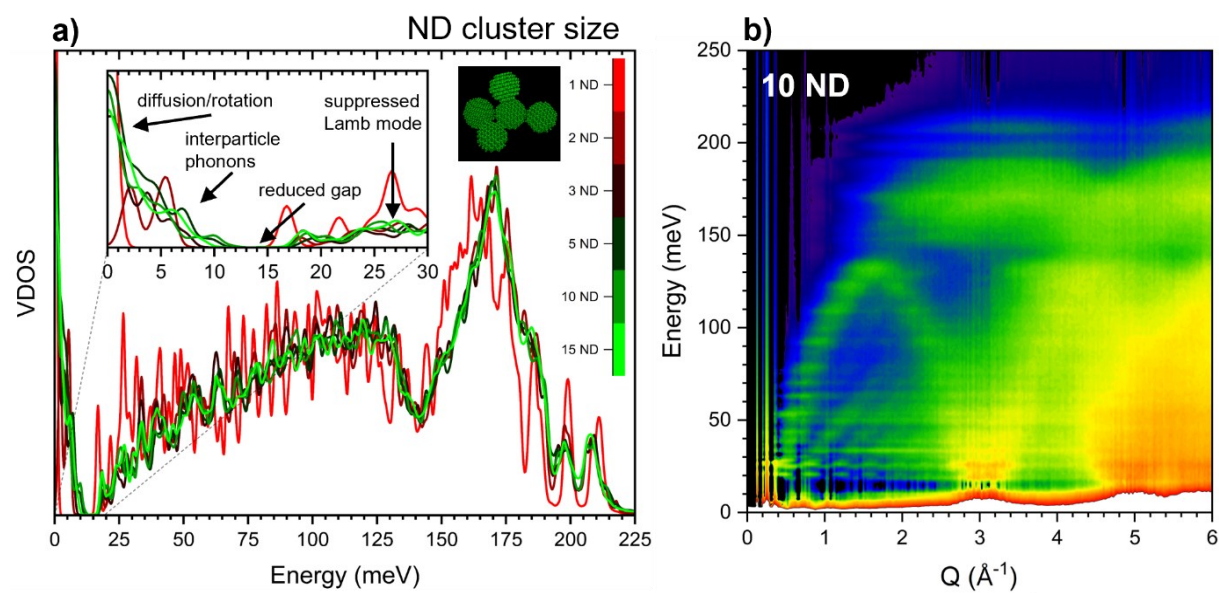

**Figure S2:** To understand the influence of inter-particle interactions on the vibrational modes of the nanocrystals, small clusters of nanodiamond were simulated. Calculated a) vibrational density of states and b) $S(Q,E)$ for a nanodiamond cluster. The DOS shows the progression of modes from the isolated ND (red) to a cluster of 15 NDs (green). The $S(Q,E)$ map is calculated for the 10 ND cluster at 300 K. The inset image shows a snapshot of an interacting ND cluster. With increasing numbers of NDs, the acoustic gap progressively closes as extended modes

can travel across particles (interparticle phonons) (inset). The Lamb resonance (~27 meV) is also suppressed with additional NDs due to interparticle coupling. In these simulations, linear and angular momentum are not zeroed during the initialisation – as they are for the other simulations in this work – so that the particles can interact freely with one another. This results in low-energy (< 2 meV) diffusive/rotational modes that lead to a non-zero DOS at E ≈ 0 (QENS) (inset).

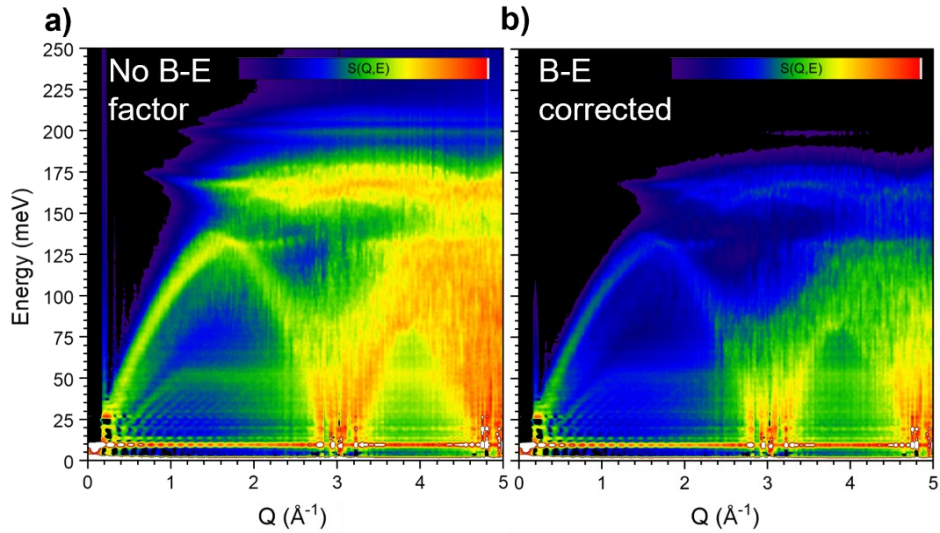

**Figure S3:** A comparison of the classical-MD-calculated $S(Q,E)$ maps for ND5 a) without a correction for the Bose-Einstein mode population ("detailed balance") (as per Fig. 1 in the main text) and b) with the Bose-Einstein correction factor. This factor is included in the calculation of the density of states for MD and INS measurements but will have some unaccounted-for effect on vibrational mode linewidths. B-E statistics dictate the scattering intensity of neutron scattering data according to detailed balance. At low energies, this effect is negligible.

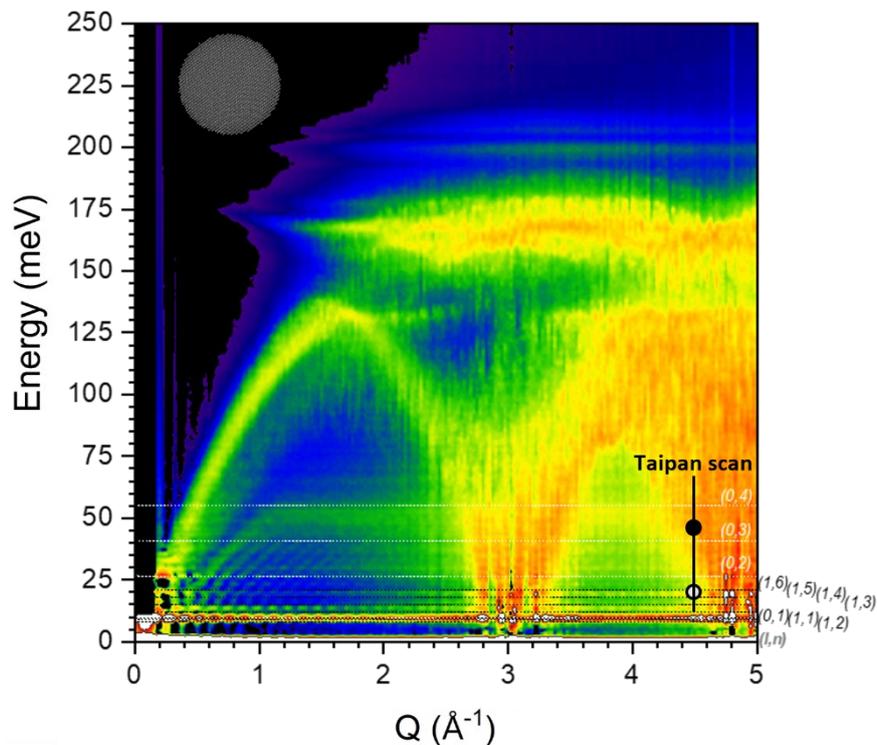

**Figure S4:** Calculated $S(Q,E)$ map for the 5.2 nm spherical ND highlighting the Lamb eigenfrequencies. Representative (and not all) Lamb resonance energies were calculated for an isotropic sphere of 5.2 nm diameter and transverse and longitudinal sound speed given by the respective phonon dispersions in the plot (~ 11 km/s

and ~ 18 km/s) and overlayed (dotted lines) with corresponding angular momentum ($l$) and overtone ($n$) numbers. The location of the scan and peaks for the experimental data taken on the triple-axis neutron spectrometer, Taipan, for the powder ND sample is also shown, where the peaks can be assigned to the TA phonon and a high-order Lamb resonance, possibly (1,5) or (1,6).

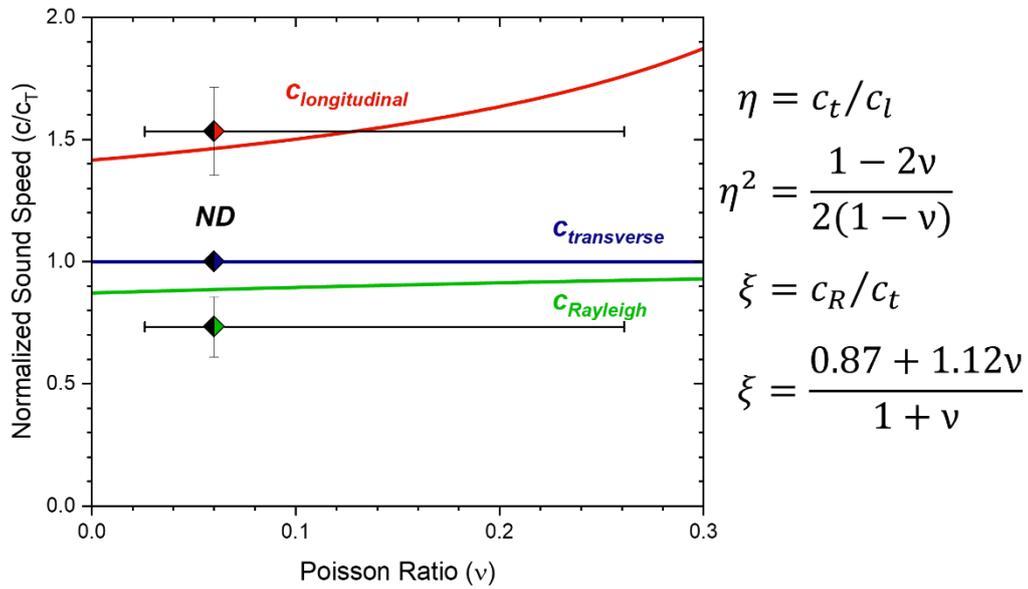

**Figure S5:** Speeds of sound for transverse, longitudinal, and Rayleigh waves, normalised to the transverse phonon calculated from the slopes of the phonon dispersions in the 5.2 nm simulated spherical nanodiamond (Fig. 1b in the main text) (y-error comes from fitting). The solid lines are calculated using the equations in the figure. The data points use the literature Poisson value for bulk diamond, while the lower x-error bound is the measured value for a "nanocrystalline" film, and the upper bound is the measured value for an "ultrananocrystalline" film (Mohr et al., *J. Appl. Phys.* **2014**, *116* (12)).

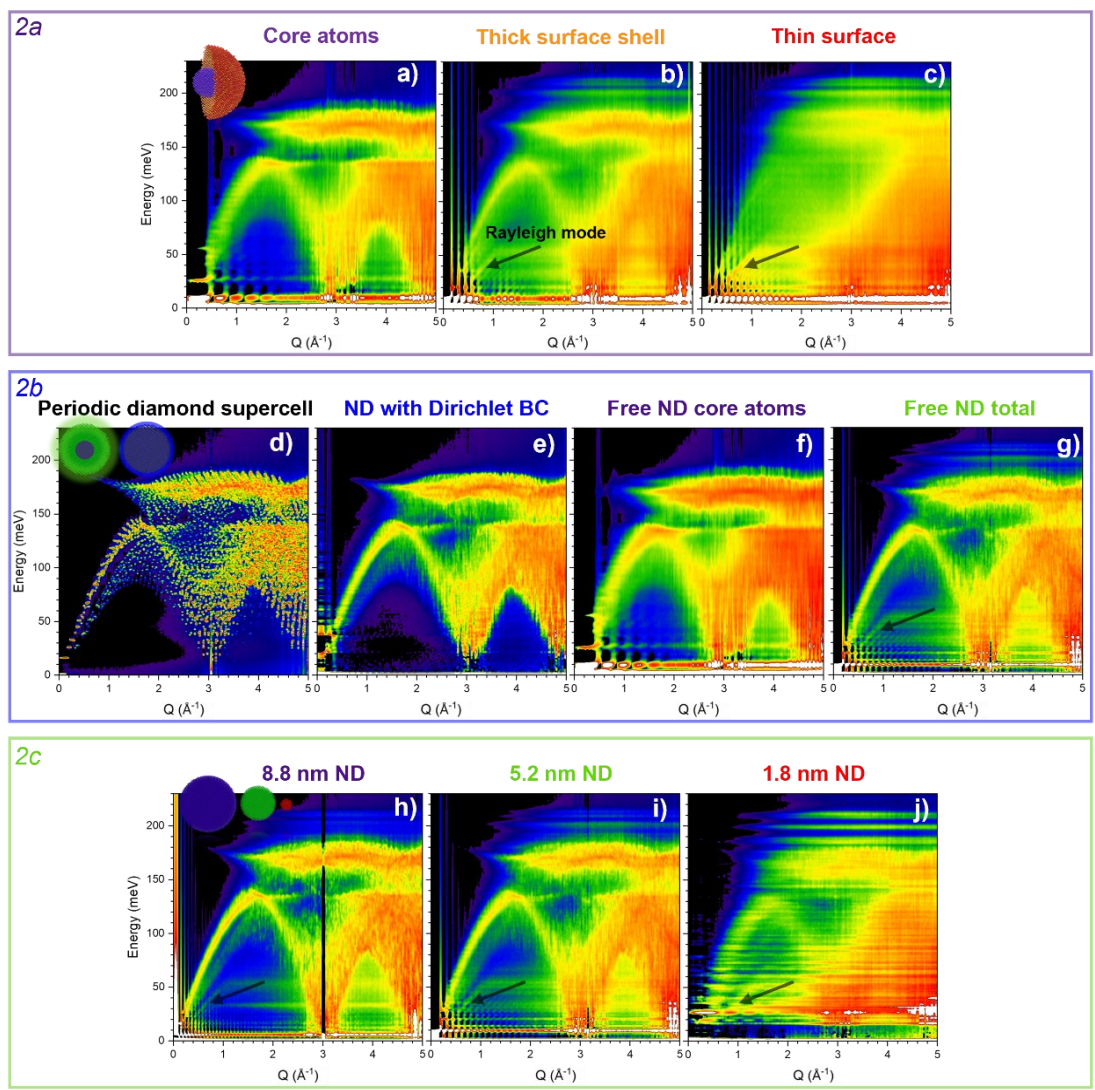

**Figure S6:** Dynamic structure factor maps corresponding to the DOS data shown in Fig. 2 in the main text. Maps a-c highlight the core-surface evolution of vibrational modes (corresponding with Fig. 2a). Maps d-g highlight the effects of rigid boundary conditions on vibrational modes (corresponding with Fig. 2b). Maps h-j highlight the particle-size evolution of the vibrational modes (corresponding to Fig. 2c). Arrows highlight the Rayleigh phonon mode which appears confined to the surface of the particle when it has free boundary conditions.

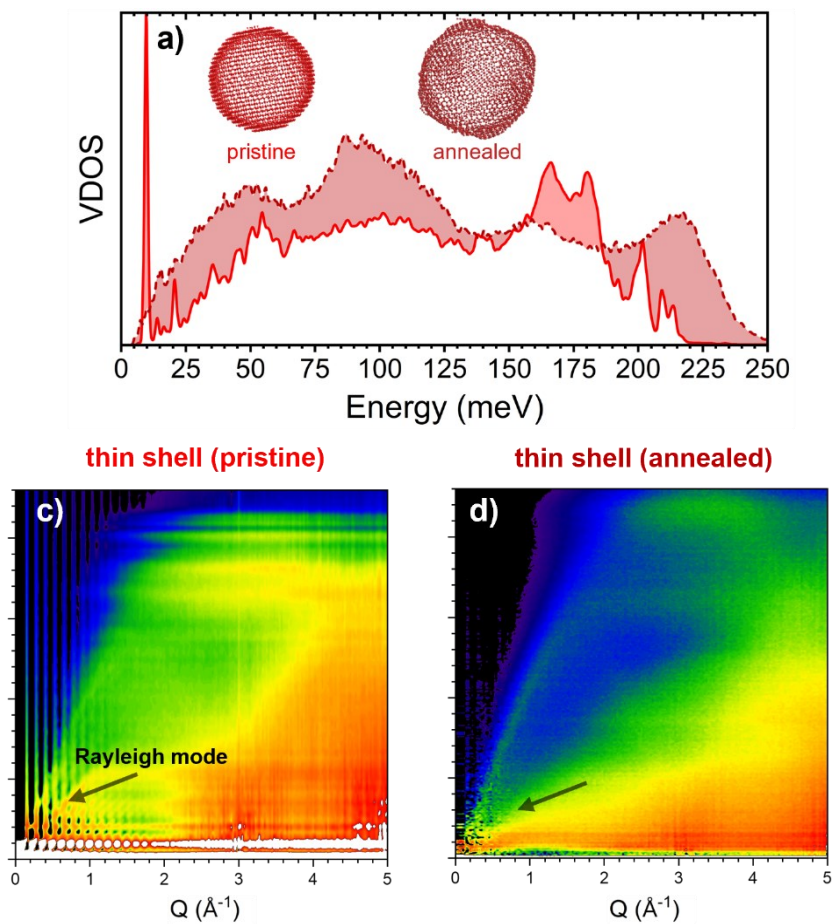

**Figure S7:** Comparison of the vibrational modes of the outer shells of the pristine and annealed nanodiamond in a) the VDOS and c), d) corresponding $S(Q,E)$ maps. The overall shape of the VDOS from the outer shell of the spherical ND and the outer shell of the annealed CS-ND are very similar, with the CS-ND shell exaggerating the features that emerge from the shell of the spherical ND. This indicates that those surface modes are enhanced by increasing disorder at the surface (i.e., they can support more of the same type of mode) or that they primarily originate from said disorder. The $S(Q,E)$ map of the CS-ND shell (d) loses almost all identifiable features as they are energy-broadened into a smeared, diffuse version of the spherical ND signal as a result of the strong structural disorder.

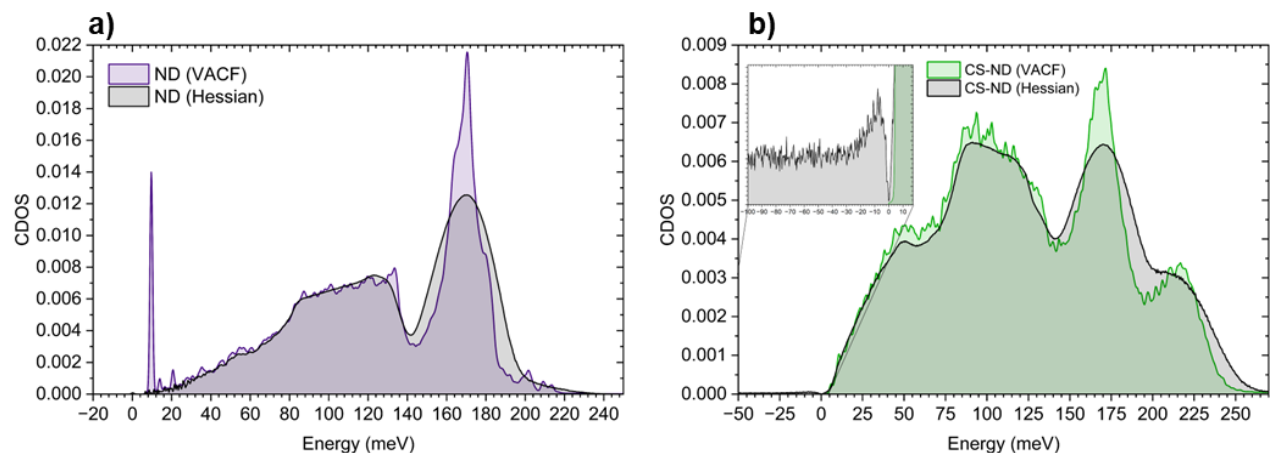

**Figure S8:** Comparison of VDOS calculated using the velocity auto-correlation function (VACF) and from the Hessian matrix (instantaneous normal mode, INM, analysis) using the same simulation outputs for the spherical 5.2 nm ND and core-shell ND. The two methods are in good agreement across all features except for the fundamental Lamb resonance (~10 meV) which is attributed to different population factors. The lack of imaginary

modes suggests that "liquid-like" motion is not the cause of the linear DOS. Nonetheless, a small number of imaginary modes are present in the CS-ND (0.7 % of the total modes - too few to significantly impact the total DOS), likely originating from the semi-amorphous shell, which should be further investigated.

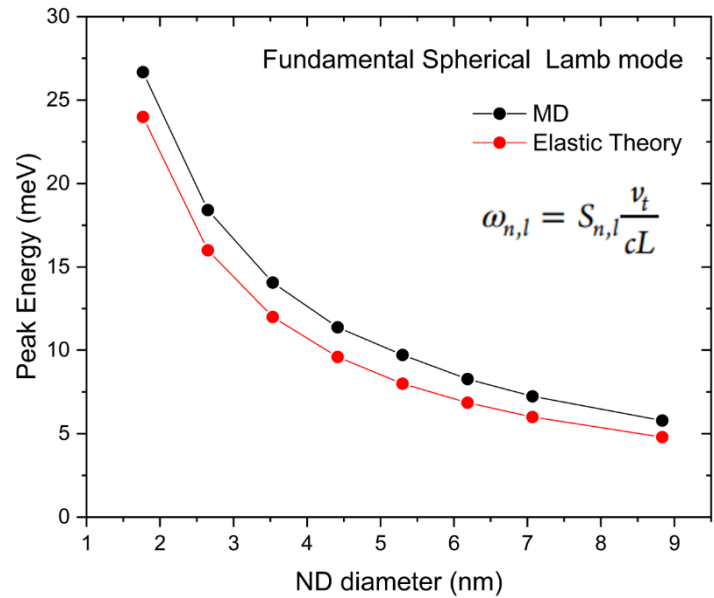

**Figure S9:** The ND size-dependence of the fundamental spherical Lamb mode as derived from the MD simulations (black) compared with the theoretical values from elasticity theory (red) using elastic values from the literature. The offset in energy highlights the overly stiff bonding in the MD simulation.

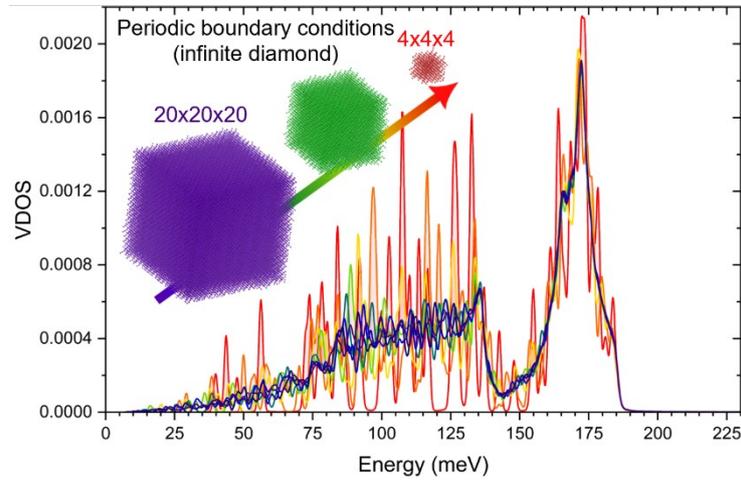

**Figure S10:** VDOS for the periodic supercells with equivalent numbers of atoms as the spherical particles in Fig. 2c in the main text, for reference.

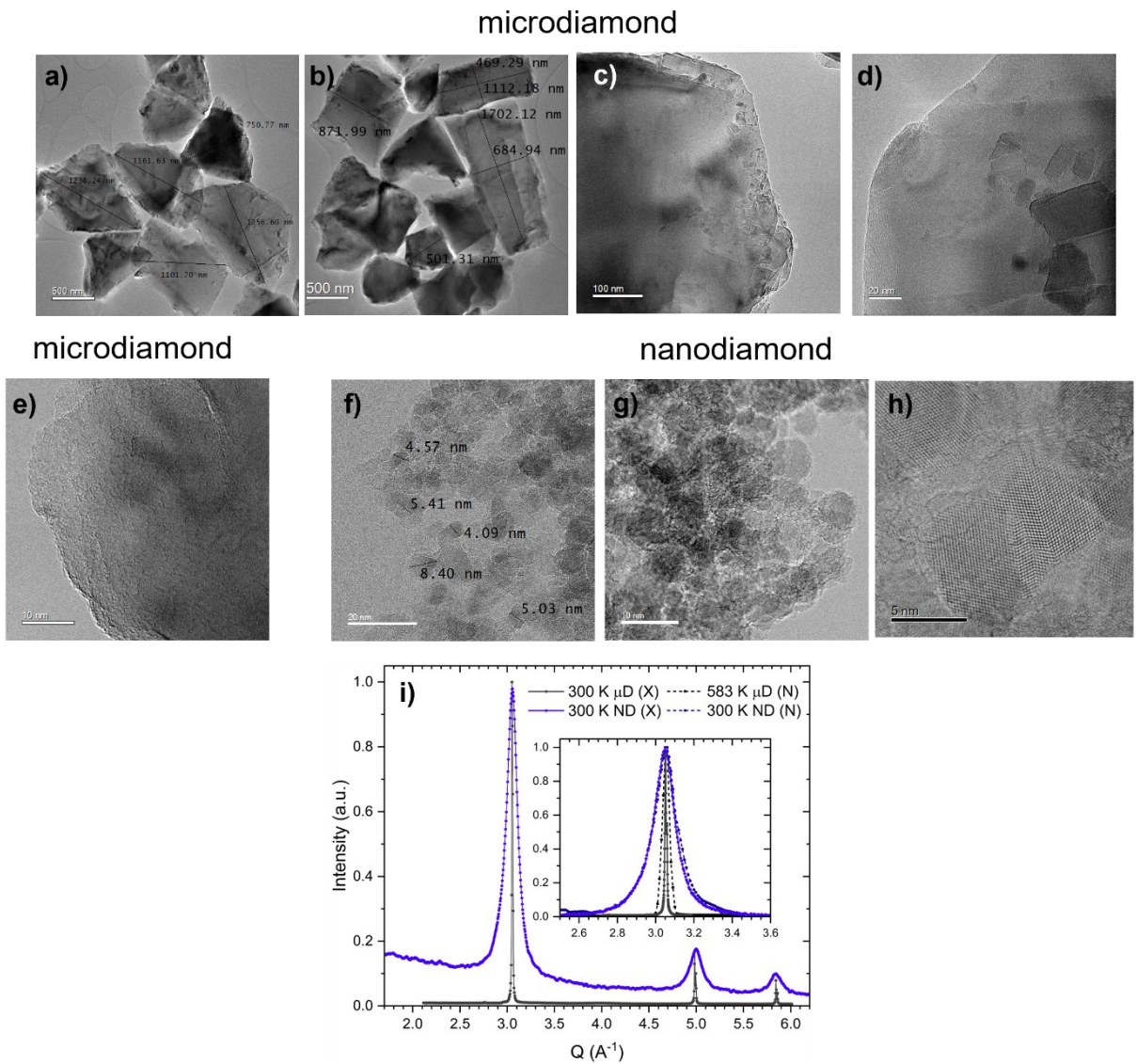

**Figure S11:** Additional TEM micrographs for a)-e) the microdiamond and f)-h) nanodiamond samples used in the neutron scattering experiments. The microdiamond samples are multi-faceted with some nano-crystallites embedded in or on the surface of the microcrystals (c-d). The edge of the μD crystals appear to have the same amorphous/graphitic layer as the NDs (d-e). The average ND size from the TEM is ~ 5 nm, with most particles in the range of 3-10 nm. i) X-ray and neutron diffraction patterns for the μD and ND samples. The average crystallite size for the ND sample from the Scherrer analysis of the XRD pattern is ~ 4 nm, which is slightly smaller than what is (on average) observed in the TEM due to the presence of the amorphous/graphitic shell.

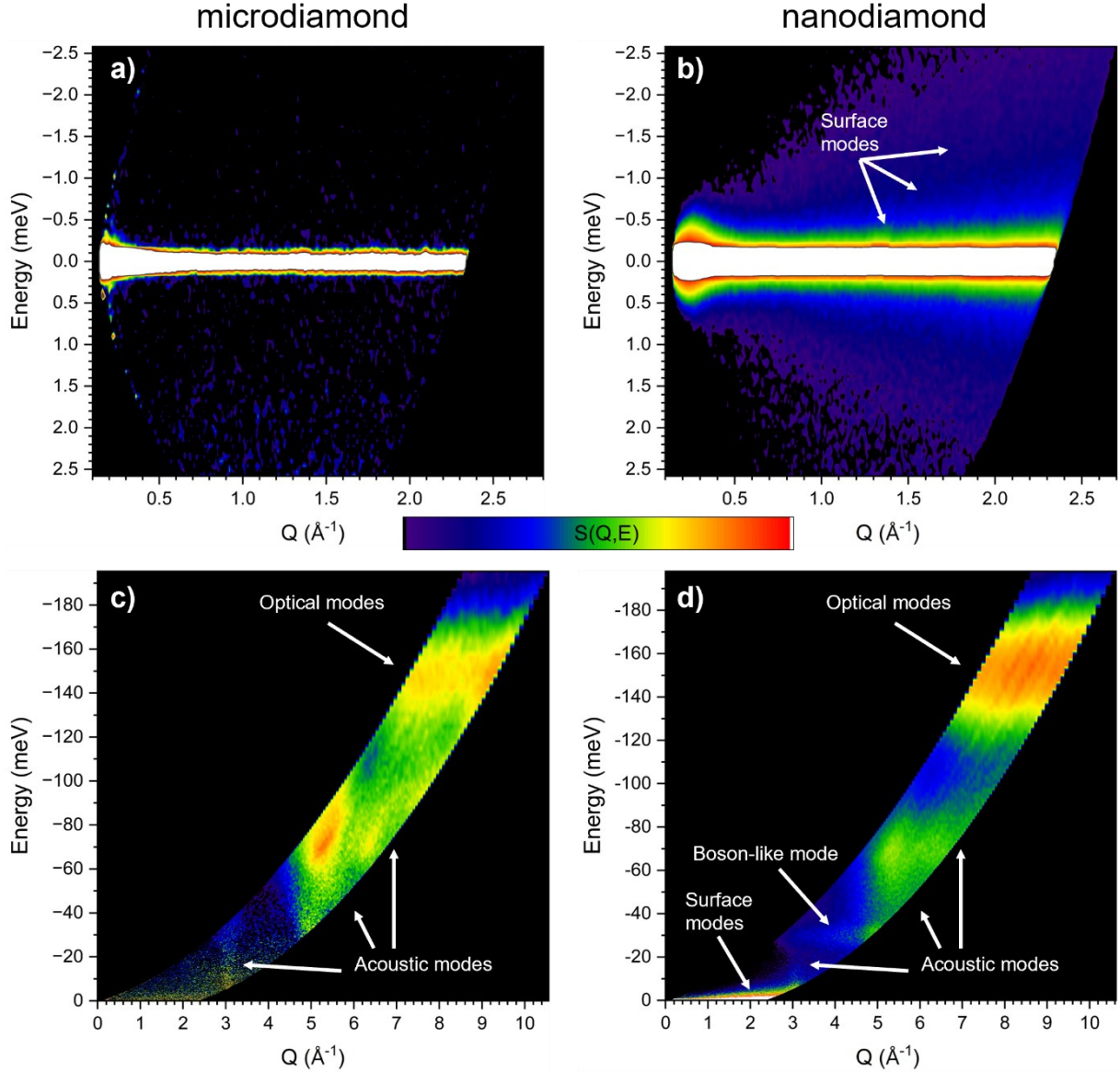

**Figure S12:** Additional INS $S(Q,E)$ maps comparing the µD (left) and ND (right) samples with an incident neutron wavelength of 4.69 Å at 583 K (main text Fig. 3f,g maps show data from incident neutron wavelength of 2.345 Å). The low-energy region across the elastic line shows a complete lack of phonon modes for a) µD, while the same region shows intense surface modes in the b) ND map. The ND map is symmetrical in features across $E = 0$ and no clear resonant modes can be distinguished on top of the surface mode signal for either neutron energy gain ($-E$) or loss ($+E$). The $S(Q,E)$ maps up to high energy show the differences in phonon modes for the c) µD and d) ND samples. The data in (c) and (d) are corrected for the Bose-Einstein phonon population at 583 K (the inverse of the correction made in Fig. S3). The broadening of core modes and the additional resonant boson-like mode can clearly be observed in the ND map. The data in (c) and (d) are used for the calculation of the VDOS in Fig. 3 in the main text (without the Bose-Einstein correction, since the DOS calculation factors that in separately).

# Tables

**Table S1:** Overview of MD systems studied. Periodic diamond (PD) supercells and nanodiamond (ND) particles of corresponding numbers of atoms were investigated. The ND5 (5.3 nm) ND particle is the primary particle of interest in this work.

| Name (PD, ND) | PD cells | PD size (length) (nm) | PD atoms | ND diameter (nm) | ND atoms |
|---|---|---|---|---|---|
| D1* | $4^3$ | 1.4 | 512 | 1.8 | 489 |
| D2 | $6^3$ | 2.1 | 1728 | 2.7 | 1755 |
| D3 | $8^3$ | 2.8 | 4096 | 3.5 | 4109 |
| D4 | $10^3$ | 3.6 | 8000 | 4.4 | 8009 |
| **D5**** | $12^3$ | 4.3 | 13824 | **5.3** | **13707** |
| D6 | $14^3$ | 5.0 | 21952 | 6.2 | 21849 |
| D7 | $16^3$ | 5.7 | 32768 | 7.1 | 32791 |
| D8 | $20^3$ | 7.1 | 64000 | 8.8 | 64097 |

*Multiple used for the ND cluster(s) (Fig. S2)

**Used to create the core-shell (CS) ND

**Table S2:** Summary of Lamb resonances investigated in nanocrystals of similar sizes to this work in the literature.

| Nanocrystal system | Size (nm) | Fundamental resonance (meV) | Method | Reference |
|---|---|---|---|---|
| ion-embedded Au | 5 | 2.8 | plasmon resonance Raman | [3] |
| isolated, spherical Au | 4.5 | unclear | MD-HD | [3] |
| ion-embedded Ag | 4.5 | 1.3 | plasmon resonance Raman | [6] |
| isolated, spherical Ag | 4.5 | unclear | MD-HD | [6] |
| diamond powder* | 2.6 | 16.2 | Raman | [7] |
| isolated, $O_h$ diamond | 2.8 | 22 | MD** | [8] |

MD = molecular dynamics, HD = hessian diagonalisation

*Bottom-up, high-pressure, high-temperature synthesis

**Vibrational mode analysis not specified, assumed to be HD